\documentstyle[12pt]{article}
\input epsf %%%  or psfig
\def\DESepsf(#1 width #2){\epsfxsize=#2 \epsfbox{#1}}
\renewcommand{\thefootnote}{\fnsymbol{footnote}}
 
\begin{document}

\begin{flushright}
          UCD-96-04\\
%          hep-ph/9601-96-04\\
        January 1996
\end{flushright}
\vspace{0.1truein}
\begin{center}
{\bf HIGGS BOSON AND \protect\boldmath$W_LW_L$ SCATTERING AT 
\protect\boldmath$e^-e^-$ COLLIDERS }\footnote{Contribution
to the Proceedings of $e^-e^-$ Workshop, Santa Cruz, CA,
Sept. 4--5, 1995.}\\
\vspace{0.37truein}
{\footnotesize TAO HAN  }\\
{\footnotesize\it Davis Institute for High Energy Physics}\\
{\footnotesize\it Department of Physics, University of California, 
Davis, CA 95616, USA}
\vspace{0.15truein}
\end{center}
\vspace{0.21truein}
\begin{abstract}
We discuss the Standard-Model Higgs boson production 
in the channels $e^-e^-\to e^-e^- H$, $e^-\nu W^- H$, and $e^-e^- ZH$.
We also illustrate the enhancements in the $W^-W^-$ cross section that 
would result from a strongly-interacting Higgs sector or from a $H^{--}$ 
resonance in a doublet + triplet scalar field model.
\end{abstract}

\section{Introduction}
\noindent
High-energy experiments at $e^+e^-$ colliders have proved 
to be very fruitful.
The construction of high energy $e^-e^-$ colliders has not been
pursued as actively. 
The probable reason for this lack of activity
is the absence of $s$-channel resonance production 
and pair production for new particles in $e^-e^-$ collisions due to
lepton number conservation. However, precisely because direct channel
resonances are not expected, high energy $e^-e^-$ collisions could be
a clean way to uncover physics beyond the Standard Model (SM). This
has become evident from the works presented in this workshop.\cite{heusch}

In this presentation, we discuss the production of the standard model 
(SM) Higgs boson ($H$), and the associated production with a weak boson.  
We also study the possibility of detecting strong $W^-W^-$ scattering 
in the $I=2$ channel, which is unique for an $e^-e^-$ collider,
and quantitatively evaluate the enhancement of $W^-W^-$ production 
due to a doubly-charged Higgs boson. The results presented here
are essentially based on two recent papers.\cite{emem,epem}

%%%%%%%%%%%%%%   PUT THIS SOMEWHERE ON THE SECOND PAGE OF TEXT %%%%%
\setcounter{footnote}{0}
\renewcommand{\thefootnote}{\alph{footnote}}
%%%%%%%%%%%%%%%%%%%%%%%%%%%%%%%%%%%%%%%%%%%%%%%%%%%%%%%%%%%%%%%%%%%%

\section{SM Higgs Boson Production}
\noindent
In $e^-e^-$ collisions single Higgs boson production at lowest order
can occur via the  processes
\begin{eqnarray}
e^-e^- \to e^-e^-H        \;, \; \ 
e^-e^- \to e^-\nu_e W^- H \;, \; \ 
e^-e^- \to e^-e^-ZH.
\label{eq:smhiggs}
\end{eqnarray}
\begin{figure}[htb]
\vspace{-1.3truein}
\centerline{ \DESepsf(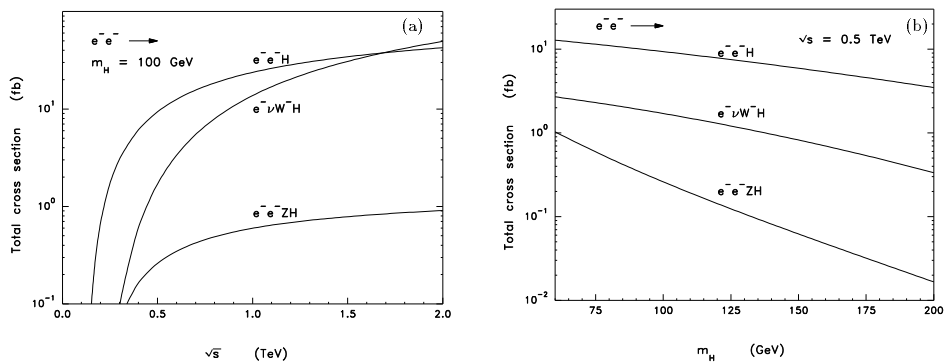 width 17 cm) }
\smallskip
\vspace{-5.2truein}
\caption {
Cross sections for production of the Standard-Model Higgs boson in
$e^-e^-$ collisions (a)~versus $\protect \sqrt s$ at $m_H=100$~GeV, 
(b)~versus $m_H$ at $\protect \sqrt s=0.5$~TeV. }
\label{fig:higgs}
\end{figure}
The cross sections are shown in Fig.~\ref{fig:higgs}(a) versus 
c. m. energy $\sqrt s$ for $m_H=100$~GeV 
and (b) versus $m_H^{}$ at $\sqrt s=0.5$~TeV. 
For $m_H^{}=100$~GeV and $\sqrt s=0.5$~TeV
the cross section is 9~fb for $e^-e^-\to e^-e^-H$ via $Z^*Z^*\to H$. 
In comparison, the Higgs production cross sections 
in $e^+e^-$ collisions for the same $\sqrt s$ and $m_H$
are 95~fb for  $W^{+*}W^{-*}\to H$ and 60~fb for $Z^*\to ZH$ 
mechanisms.\cite{bckp}
Searching for the SM Higgs boson in $e^-e^-$ collisions
is limited by the rather small cross sections, especially
for heavier $m_H^{}$. 

For $m_H^{} < 150$ GeV or so, 
the Higgs boson decays dominantly to $b\bar b$ and 
a Higgs signal identification is quite feasible.
The only complication 
comes from the case in which $m_H$ is close to $M_Z$,
when the $Z$-production processes 
\begin{eqnarray}
e^-e^- \to e^- e^-Z \;, \; \ 
e^-e^- \to e^-\nu_e W^-Z \;, \; \ 
e^-e^- \to e^-e^- ZZ
\label{eq:hback}
\end{eqnarray}
are large backgrounds to the Higgs processes (\ref{eq:smhiggs})
respectively. For example, for the production of a 100~GeV Higgs
boson at $\sqrt s=0.5$~TeV, 
$\sigma(e^-e^-\to e^-e^-H \to e^-e^- b\bar b)\approx 9$~fb and
$\sigma(e^-e^-\to e^-e^-Z\to e^-e^- b\bar b)\approx 1100$~fb
(with $B(Z\to b\bar b) = 15.5\%$). However, it is important to 
notice that the Higgs signal is very distinctive.
The $H$-production is central while the $Z$-production is peaked at
forward and backward scattering angles; thus the $Z$ background can be
selectively suppressed by angular cuts.  
\begin{figure}[htb]
\vspace{-1.65truein}
\centerline{ \DESepsf(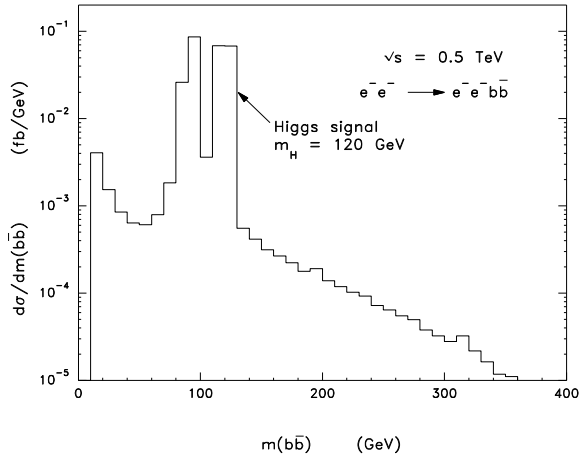 width 11 cm) }
\smallskip
\vspace{-1.3truein}
\caption {
The differential cross section 
$d\protect \sigma/d m(b \protect \bar b)$ versus the
invariant mass $m(b\protect \bar b)$ of the 
$b\protect \bar b$ pair in the process
%$e^-e^- \to e^- e^- b\protect \bar b$ at $\protect \sqrt s=0.5$~TeV,
with the acceptance cuts of
Eqs.~(\protect \ref{pTe cut}) and (\protect \ref{pTb cut}).  
The peak at $m(b\protect \bar b)\protect \approx M_Z$ 
is due to $e^-e^-\to e^-e^- Z$ with $Z\to b \protect \bar b$. 
The signal due to a Higgs boson of mass $m_H=120$~GeV is illustrated.
}
\label{fig:mhiggs}
\end{figure}

To address the $H$ signal observability quantitatively, we
evaluate the complete $e^-e^-\to e^- e^- b \bar b$ background 
including both
$e^-e^-\to e^-e^-Z$ with $Z\to b\bar b$ 
and the two-photon production of $b\bar b$.  
Based on the fact that the two-photon background can be
reduced substantially by keeping the photon propagators far off-shell,
we impose the following acceptance cuts
\begin{equation}
p_{Te} > 15\;{\rm GeV} \qquad {\rm and} \qquad
|\cos\theta_e| < \cos (15^\circ) \;, \label{pTe cut}
\end{equation}
on both of the electrons in the final state.  We also impose the
following acceptance cuts on the $b$'s in the final state
(since the $b$'s in the signal events are populated
at high transverse momenta $\approx m_H/2$):
\begin{equation}
p_{Tb} > 25\;{\rm GeV} \qquad {\rm and} \qquad  |\cos\theta_b| < 0.7 \;.
\label{pTb cut}
\end{equation}
\begin{table}
\centering
\caption{
The Higgs boson signal in the production of $e^-e^-\to e^-e^- H$ for
$m_H=60$--$140$ GeV and the background from $e^-e^-\to e^-e^- b\bar b$
with the invariant mass of the $b\bar b$ pair in the range $m_H\pm
\Delta m_H$, with $\Delta m_H=10$~GeV.  For simplicity we take B$(H\to
b\bar b)=1$, and assume that all the signal falls within $m_H\pm\Delta m_H$.
The acceptance cuts are $p_{Te} > 15$ GeV and $|\cos\theta_e| <
\cos (15^\circ)$ on the final state electrons, and $p_{Tb} > 25$ GeV
and $|\cos\theta_b| < 0.7$ on the final state $b$'s.  }
\label{table1}
\medskip
\begin{tabular}{|c|c|c|}
\hline
$m_H$     &   Signal (fb)   &  Background (fb) \\
\hline
\hline
60        &     1.4         & 0.02  \\
70        &     1.4         & 0.03  \\
80        &     1.4         & 0.27   \\
90        &     1.4         & 1.1    \\
100       &     1.4         & 0.86    \\
120       &     1.3         & 0.02    \\
140       &     1.3         & 0.01   \\
\hline
\end{tabular}
\end{table}
After imposing (\ref{pTe cut}) and (\ref{pTb cut}), the Higgs signal is
1.4~fb for $m_H=100$~GeV while the total $eeb\bar b$ background
integrated over all $m(b\bar b)$ is 1.3~fb.  The background has a
wide $m(b\bar b)$ invariant mass distribution (see
Fig.~\ref{fig:mhiggs}) with a peak at $m(b\bar b)=M_Z$ due to $e^-e^-\to
e^-e^-Z$.  The signal is a sharp peak at the Higgs mass.  We consider
an invariant mass resolution $\Delta m_H$ for the $b\bar b$ pair of 10
GeV and assume that all the signal falls within $m_H\pm \Delta m_H$.
The signal and the background in such bins at various Higgs mass
values are summarized in Table~\ref{table1}.  In this simple
comparison we have assumed perfect $b$-tagging with only the $b\bar b$
final state counted as background, {\it i.e.}, the other $q\bar
q\;(q=u,d,s,c)$ final states are rejected.  The background for
$m_H=90$ GeV is the largest because of $ee\to eeZ$.  We conclude
that the intermediate mass Higgs boson in the channel $e^-e^-\to e^-
e^- H$ with $H\to b\bar b$ may be observable with an integrated
luminosity of 20 fb$^{-1}$.

When $m_H>2m_Z$,  Higgs
production with $H\to ZZ$ decay 
can give an appreciable enhancement to
the $e^-e^-\to e^-e^-ZZ$ production; for example, at $m_H=200$~GeV the
cross section of $e^-e^-\to e^-e^-ZZ$ is a factor of two larger than
that with $m_H=100$~GeV.

\section{Strong \protect\boldmath$W_L^- W_L^-$  Scattering Signal}
\noindent
If no light Higgs boson is found for $m_H^{}$ to be less than 
about 800 GeV,  one would
anticipate that the interactions among longitudinal vector bosons
become strong.\cite{chan-gail} 
An $e^-e^-$ collider offers a unique
opportunity to explore the weak isospin $I=2$ 
$s$-channel\cite{wpwp}
via the process $W_L^-W_L^- \to W_L^-W_L^-$.\cite{hantalk}  
The
simplest model for a strongly-interacting $W_L^-W_L^-$ sector is the
exchange of a heavy Higgs boson. This results in an enhancement of the
$e^-e^-\to\nu\nu W^-W^-$ production cross section compared to that
expected from the exchange of a light Higgs boson. This enhancement
due to a Higgs boson of mass 1~TeV can be defined as the difference of
the $W_L^-W_L^-\to W_L^-W_L^-$ fusion contributions
\begin{equation}
\Delta\sigma_H = \sigma(m_H=1~{\rm TeV}) - \sigma(m_H=0.1~\rm TeV)
\end{equation}
to $e^-e^-\to\nu\nu W^-W^-$ production. There is no appreciable
numerical change between the choices $m_H=0.1$~TeV and $m_H=0$ for the
light Higgs boson reference mass. We find the values
\begin{equation}
 \Delta \sigma_H \simeq \left\{
\begin{array}{ll}
53.6 - 50.9 = 2.7~{\rm fb}
& \mbox{at $\sqrt s = 1.5$~{\rm TeV}; } \\
86.5 - 82.0 = 4.5 ~{\rm fb}
& \mbox{at $\sqrt s = 2$~{\rm TeV}. }
\end{array}
\right.
\label{eq:rate}
\end{equation}

We first address the observability of this strong
$W_L^-W_L^-\to W_L^-W_L^-$ signal at $\sqrt s=2$~TeV. 
The cross section for $e^-e^-\to\nu\nu W^-W^-$ from all Standard-Model
diagrams (dominantly $W_T^- W_T^-$) is about 20 times larger than
$\Delta\sigma$. Hence the background contributions associated with
transverse $W$ bosons ($W_T^-W_T^-$, $W_T^-W_L^-$) must somehow be
selectively reduced by acceptance criteria if we are to observe the
strongly-interacting $W_L^-W_L^-$ signal.

There are several ways to accomplish the substantial
background suppression.\cite{wpwp,hantalk,baggeretal}  The
$W_L^-W_L^-$ scattering process gives large $M(W^-W^-)$ of order 1~TeV
with centrally-produced $W^-$ having large $p_T$. Thus we impose the
kinematic cuts
\begin{eqnarray}
 p_T(W) > 150 ~{\rm GeV},  \quad  |\cos \theta_W| < 0.8 \label{pT(W)} \,,
\end{eqnarray}
which retains about one-third of the signal and reduces the SM
backgrounds by more than an order of magnitude.  After those cuts, the
heavy Higgs enhancement becomes 
$\Delta \sigma_H \simeq 7.8 - 6.3 = 1.5$~fb.

In hadronic $W$-decays, the sign of the $W$ charge is not identified
and the two-photon process $e^-e^-\to e^-e^-W^+W^-$ may also present a
substantial background when the final electrons are not observed.
The cross section for $e^-e^-\to e^-e^-W^+W^-$
is about a factor of 30 larger than $e^-e^-\to \nu\nu
W^-W^-$. Moreover, $e^-e^-\to e^-\nu W^-Z$ could add to the background
if the reconstructed invariant masses in hadronic decays are not
sufficiently resolved to distinguish $W$ from $Z$. In order to
suppress these backgrounds we veto events in which an electron can be
identified having
\begin{eqnarray}
 E_e > 50 ~{\rm GeV},  \quad  |\cos \theta_e| < |\cos(150~{\rm mrad})| \,.
\label{E_e}
\end{eqnarray}
With the acceptance of Eqs.~(\ref{pT(W)}) and (\ref{E_e}), the
remaining $e^-e^-\to e^-e^-W^+W^-$ and $e^-e^-\to e^-\nu_e W^-Z$
backgrounds are 60 fb and 10 fb, respectively.

\begin{figure}[htb]
%\begin{figure}[t]
\vspace{-1.2truein}
\centerline{ \DESepsf(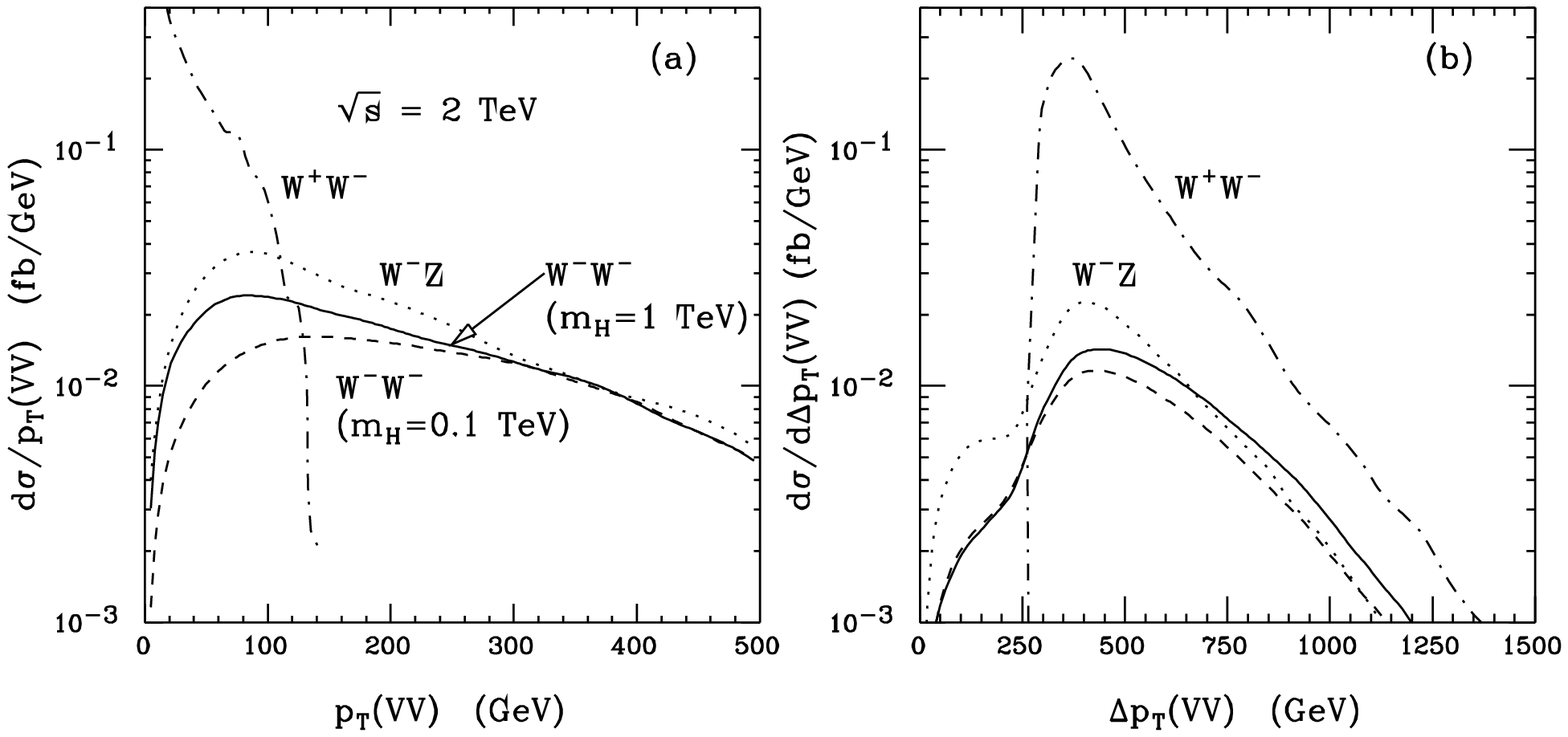 width 11 cm) }
\smallskip
\vspace{-2.2truein}
\caption {
Distributions at $\protect \sqrt s=2$~TeV in the transverse
momenta of vector bosons produced in the reactions $e^-e^-\to\nu\nu
W^-W^-, e^-e^-W^+W^-, e^-\nu W^-Z$: 
(a)~$p_T(VV) = \left|{\bf p}_T(V_1)+{\bf p}_T(V_2)\right|$, 
(b)~$\Delta p_T(VV) = \left|{\bf p}_T(V_1)-{\bf p}_T(V_2)\right|$.
}
\label{fig:ptvv}
\end{figure}

A further improvement in isolating the $W_L^-W_L^-$ signal derives
from the fact that the $p_T(WW)$ spectrum of the signal is peaked
around $M_W$ and falls off rapidly at high $p_T$ like $1/p_T^4$.
Figure~\ref{fig:ptvv}(a) compares the $p_T(VV)$ distribution of the
$W_L^-W_L^-$ signal with the backgrounds. Note that the difference
between the solid curve (with $m_H=1$~TeV) and the dashed curve
(with $m_H=0.1$~TeV) is the strong $W_L^-W_L^-$ enhancement. We impose
the selection
\begin{equation}
50~{\rm GeV} < p_T(VV) < 300~\rm GeV \label{pT(VV)}
\end{equation}
for additional background suppression.

\begin{figure}[htb]
%\begin{figure}[t]
\vspace{-1.2truein}
\centerline{ \DESepsf(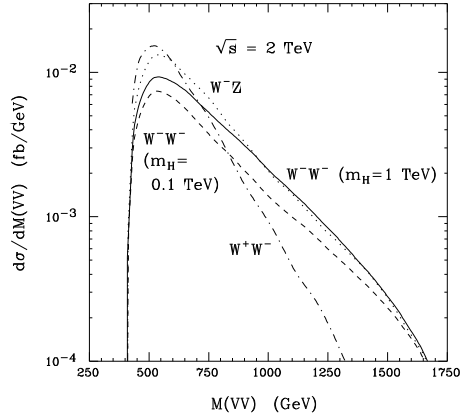 width 11 cm) }
\smallskip
\vspace{-2.0truein}
\caption {
Invariant mass distributions of the weak boson
pairs produced in the reactions $e^-e^-\to\nu\nu W^-W^-, e^-e^-W^+W^-,
e^-\nu W^-Z$ after the acceptance cuts of Eqs.~(\protect \ref{pT(VV)}) 
and (\protect \ref{Delta pT}) have been applied to enhance 
the strongly-interacting
$W^-W^-$ signal due to the exchange of a 1~TeV SM Higgs boson.
}
\label{fig:mvv}
\end{figure}

The signal gives $W_L^-$ bosons that are fast and moving back-to-back
in the transverse plane. The difference in the transverse momenta of
the two weak bosons is
\begin{equation}
\Delta p_T(VV) = |{\bf p}_T(V_1)-{\bf p}_T(V_2)|
\end{equation}
presented in Fig.~\ref{fig:ptvv}(b). The signal (difference of solid and
dashed curves) is enhanced by the cut
\begin{equation}
\Delta p_T (VV) > 400\ \rm GeV \,. \label{Delta pT}
\end{equation}
With the additional cuts of Eqs.~(\ref{pT(VV)}) and (\ref{Delta pT})
the surviving signal is
\begin{equation}
\Delta \sigma_H \simeq 3.8 - 2.8 = 1.0 \ \rm fb \,.
\end{equation}
The efficiency for retaining the signal with such cuts is 67\%.
The remaining backgrounds are 4.4~fb for $e^-e^-\to e^-e^- W^-W^+$ 
and 4.7~fb for $e^-e^-\to e^-\nu W^-Z$.
The resulting $M(VV)$ distributions after these cuts are presented in
Fig.~\ref{fig:mvv}. At high $VV$ invariant masses the strong
$W_L^-W_L^-$ scattering rate due to the exchange of a 1~TeV Higgs
boson is enhanced over the $W^-_TW^-_T$, $W_T^-W_L^-$ and $W^-W^+$ backgrounds,
while the background due to $W^-Z$ still persists.

Next we estimate the signal rates for other strongly-interacting
scenarios.  We consider a chirally-coupled scalar boson ($m_S=1$ TeV
and $\Gamma_S=350$ GeV), a chirally-coupled vector boson ($m_V=1$ TeV
and $\Gamma_V=25$ GeV),\cite{hantalk,baggeretal} and the low energy
theorem amplitude.\cite{baggeretal} These calculations are
carried out with the effective $W$-boson approximation (EWA).
In this calculational method one is unable to obtain the
exact kinematics for the final state of $W^-W^-$.  In order to
simulate the acceptance effects of Eqs.~(\ref{pT(VV)}) and (\ref{Delta
pT}), we multiplied the EWA calculations by the efficiency factor 67\%
found in the heavy Higgs boson model.
\begin{table}[htb]
\centering
\caption{
Signals at $\protect \sqrt s=2$~TeV from different
models of strongly-interacting $W^-W^-$ with cuts discussed in the
text. Backgrounds are summed over $W^-W^-$ with a light Higgs
exchange, $W^+W^-$, and $W^-Z$. 
Entries correspond to the number of events with hadronic $W,Z$
decays for an integrated luminosity of 300 fb$^{-1}$
and those in parentheses are in units of fb without the branching
fractions. As a rough indication of the signal observability, 
values of $S/\protect \sqrt B$ are also given.}
\label{table2}
\medskip
\setlength{\tabcolsep}{5pt}
\begin{tabular}{|l|c|c|c|c|c|c|} \hline
$\protect \sqrt s=2$ TeV & SM  & Scalar & Vector   & LET & Bckgnds \\
$M_{WW}^{min}$ & $m_H=1$ TeV & $m_S=1$ TeV & $m_V=1$ TeV & & \\
\hline\hline
0.5 TeV
& 130 (0.88)  & 175 (1.2)  & 167 (1.1) & 245 (1.7) & 1470 (10) \\
$S/\protect \sqrt B$
& 3.4  & 4.6  & 4.4 & 6.4 & {} \\ \hline
0.75 TeV
&  65 (0.44) & 106 (0.72) & 93 (0.63) & 150 (1.0)  & 515 (3.5) \\
$S/\protect \sqrt B$
& 2.9  &  4.7 & 4.1 & 6.6 & {} \\ \hline
\end{tabular}
\end{table}

The predicted cross sections at $\sqrt s = 2$~TeV with the cuts discussed
above are presented in Table~\ref{table2}. 
The number of events with hadronic $W,Z$ decays 
are given for an integrated
luminosity of 300~fb$^{-1}$.  In Table~\ref{table2} we see that the
backgrounds (dominantly from $W^-Z$ production) to the signals are
still substantial after the kinematic selection criteria.  Due to the
absence of an $s$-channel resonance, the signals are mostly an overall
enhancement on the $M_{WW}$ spectrum.  If we can predict the SM
backgrounds at a level of better than 10\%, there is a chance that we
can observe the strong $W^-W^-$ scattering 
via the hadronic decay modes at statistical significance
$S/\sqrt B > 4$ for a 1~TeV scalar or a vector particle, and at
$S/\sqrt B \geq 6.4$ for the LET amplitude 
with $M_{WW} > 500$~GeV.

\begin{table}[htb]
\centering
\caption{
Signals at $\protect \sqrt s=1.5$~TeV from different
models of strongly-interacting $W^-W^-$ with cuts discussed in the
text. Backgrounds are summed over $W^-W^-$ with a light Higgs
exchange, $W^+W^-$, and $W^-Z$. 
Entries correspond to the number of events with hadronic $W,Z$
decays for an integrated luminosity of 300 fb$^{-1}$. 
$W/Z$ identification via dijet mass has been implemented,
as discussed in
the text to improve the signal/background ratio.
As a rough indication of the signal observability, 
values of $S/\protect \sqrt B$ are also given.}
\label{table3}
\medskip
\begin{tabular}{|l|c|c|c|c|c|c|} \hline
$\protect \sqrt s = 1.5$ TeV & SM  & Scalar & Vector   & LET & Bckgnds \\
$M_{WW}^{min}$ & $m_H=1$ TeV & $m_S=1$ TeV & $m_V=1$ TeV & & \\
\hline\hline
0.5 TeV
& 41  & 53   & 54 & 63  & 345  \\
$S/\protect \sqrt B$
& 2.2  & 2.8  & 2.9 & 3.4 & {} \\ \hline
\end{tabular}
\end{table}

At $\sqrt s=1.5$ TeV, the signal rate is reduced
by 40\%, as shown in Eq.~(\ref{eq:rate}), which makes
the signal observation more difficult. An improvement 
was made to include the $W/Z$ discrimination through
the di-jet mass of their decay products.
Typically, one assumes the jet energy resolution to be\cite{jlc}
\begin{eqnarray}
\delta E_j/E_j = 0.50 \Big/ \sqrt{E_j} \;\oplus\; 0.02 \;,
\end{eqnarray}
in GeV units (where the symbol $\oplus$ means adding in quadrature).
If we now identify dijets having measured mass in the intervals
$$\left[0.85M_W, \; {1\over 2}(M_W+M_Z)\right] \quad
{\rm and} \quad \left[{1\over 2}(M_W+M_Z), \; 1.15M_Z\right]$$
as $W \to jj$ and $Z\to jj$, respectively, then simulation\cite{epem}
indicates that true $W W$, $W Z$, $ZZ\to jjjj$ 
events will be interpreted statistically as follows:
$$
\begin{array}{lcrrrrr}
WW &\Rightarrow 
& 73\%\: WW, & 17\%\: WZ, &  1\%\: ZZ,&  9\%\: {\rm reject},
\\
WZ &\Rightarrow 
& 19\%\: WW, & 66\%\: WZ, & 7\%\: ZZ,&  8\%\: {\rm reject},
\\
ZZ &\Rightarrow 
&  5\%\: WW, & 32\%\: WZ, & 55\%\: ZZ,&  8\%\: {\rm reject}.
\end{array}
$$
This helps improve the signal observability and the results
are shown in Table~\ref{table3}.

The use of $Z \to e^+ e^-, \mu^+\mu^-, b \bar b$ (with
$b$-tagging), 
with combined branching fraction of about 22\%, could be
helpful in determining the contribution of the $W^-Z$ 
background process and further improving the signal identification.

\section{\protect\boldmath$H^{--}\rightarrow W^-W^-$ 
Signal in Scalar Doublet + Triplet Model}

\begin{figure}[htb]
%\begin{figure}[t]
\vspace{-1.2truein}
\centerline{ \DESepsf(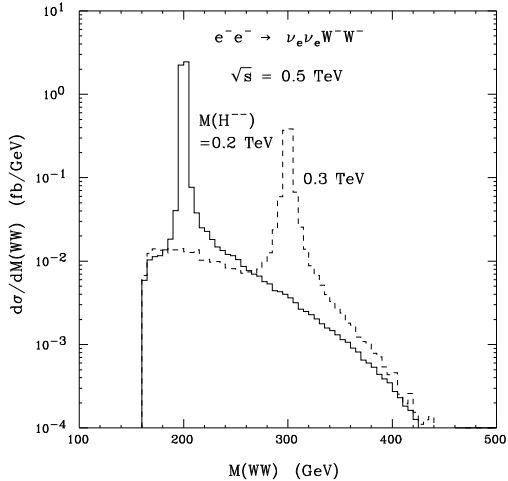 width 11 cm) }
\smallskip
\vspace{-1.8truein}
\caption {
The distribution in the $W^-W^-$ invariant mass for $e^-e^-\to
\nu_e\nu_e W^-W^-$ including the contribution of a doubly-charged
Higgs boson of mass $M(H^{--})=0.2$ or 0.3~TeV.
}
\label{fig:hmm}
\end{figure}

\noindent
A search for a doubly-charged Higgs boson in a model with a scalar
triplet could also be carried out at an $e^-e^-$ collider. 
 What makes
the $e^-e^-$ collider unique in this instance is that the
doubly-charged Higgs boson can be produced as an $s$-channel resonance.
Jack Gunion\cite{gunion1} recently discussed the lepton-number
violating process
$$
e^- e^- \rightarrow H^{--}
$$
and found interesting results.  We here consider the process
$$
W^-W^- \to H^{--}
$$ 
followed by $H^{--}\to W^-W^-$ decays.
Detailed analyses of the doublet
plus triplet model can be found in the 
literature\cite{gunion2} and we
will not repeat the discussion of the model here.
With certain assumptions for simplicity,
there are then two independent parameters left in the 
model:\cite{emem,gunion2} the mass
parameter $M(H^{--})$ and the mixing angle $\theta_H$
between the Higgs doublet and the triplet fields, which is related to
the ratio of the vacuum expectation values.
Figure~\ref{fig:hmm} presents the $M_{W^-W^-}$ distribution at
$\sqrt s=0.5$~TeV including the resonance process $e^-e^- \to
\nu \nu H^{--} \to \nu \nu W^-W^-$ with $M(H^{--})=0.2$ or
0.3~GeV, taking maximum mixing $\tan\theta_H=1$.  A significant
enhancement above the SM background occurs. If existed at all, 
such a doubly-charged Higgs boson should manifest itself in
$e^- e^-$ collisions.

Discussions here should be essentially applicable for
the doubly-charged vector bosons as well.\cite{frampton}

\section{Summary}
\noindent
We have calculated the single and associated Higgs boson production
and found that they may give observable signals, 
although the cross sections are generally not
as large as those in $e^+e^-$ collisions.  We have also investigated
the possibility of observing strong $W^-_LW^-_L$ scattering, which
occurs through the weak isospin $I=2$ channel and is unique to
$e^-e^-$ collisions. We have developed certain kinematic cuts to
significantly reduce the $W^-_TW^-_T$, $W_T^-W_L^-$, $W^+W^-$ and $W^-Z$ backgrounds to the strong $W_L^- W^-_L$ signal in hadronic decay modes.
The $W^-Z$ background persists at
large $M_{WW}$, which makes the observation of strong $W^-_LW^-_L$
scattering difficult. On the other hand, if good jet energy resolution
can be obtained, then $W$ and $Z$ may be distinguished so that the
``faked'' background from $WZ$ final state may be further suppressed.
If doubly-charged Higgs bosons exist, 
the $s$-channel enhancement in $W^-W^-$ final state
would be very substantial at an $e^-e^-$ collider. 
Finally, by colliding $e_L^-$
beams the strong $W^-_L W^-_L$ signal in the process $e^-e^-\to \nu_e
\nu_e W^-_L W^-_L$ can be enhanced over the $W^+W^-$, $ZZ$, and $W^-Z$
backgrounds.  Similarly, the $H^{--}$ signal 
will be favored with the use of $e^-_L$ beams.
This survey of
cross sections and processes should provide useful benchmarks for
serious studies of the potential of such a machine for new physics
discovery. 

\vskip1.5pc
\leftline{\bf Acknowledgments}
 
\vskip6pt\noindent
I thank V. Barger, J. Beacom, K. Cheung and R. Phillips for 
the collaboration that led to most of the results presented here.
This work is supported in part by the U.S. Department of Energy 
under contract DE-FG03-91ER40674, and by the Davis Institute
for High Energy Physics.

%%%%%%%%%%%%%%%%%%%%%%%%%%%%%%%%%%%%%%%%%%%%%%%%%%%%%%%%%%%%%%%%%

\end{document}